\documentclass[12pt]{cernart}
\usepackage{epsfig}
\usepackage{graphicx} 
\usepackage{amsmath,amssymb}
\usepackage{amssymb}

\begin{document}

\begin{titlepage}
\docnum{CERN-PH-EP/2005-037}
\date{13~July~2005}

\begin{center}
\title{A measurement of the CP-conserving component 
of the decay $K_S^0 \rightarrow \pi^+ \pi^- \pi^0$ \\
}
% \author{The NA48 collaboration}
\begin{Authlist}
\begin{center}
%{\bf Experiment NA48/1}
\  \\[0.2cm] 
% A.~Lai,
% D.~Marras \\
%{\em \small Dipartimento di Fisica dell'Universit\`a e Sezione dell'INFN di Cagliari, I-09100 Cagliari, Italy} \\[0.2cm] 
 % 
%
J.R.~Batley,
%         G.E.~Kalmus,
C.~Lazzeroni,
D.J.~Munday,
M.~Patel\footnotemark[1],
M.W.~Slater,
S.A.~Wotton \\
{\em \small Cavendish Laboratory, University of Cambridge, Cambridge, CB3 0HE, U.K.\footnotemark[2]} \\[0.2cm] 
 R.~Arcidiacono,
 G.~Bocquet,
 A.~Ceccucci,
 D.~Cundy\footnotemark[3],
 N.~Doble,
 V.~Falaleev,
 L.~Gatignon,
 A.~Gonidec,
 P.~Grafstr\"om,
 W.~Kubischta,
F.~Marchetto\footnotemark[4],
 I.~Mikulec\footnotemark[5],
 A.~Norton,
 B.~Panzer-Steindel,
P.~Rubin\footnotemark[6],
 H.~Wahl\footnotemark[7] \\
{\em \small CERN, CH-1211 Gen\`eve 23, Switzerland} \\[0.2cm] 
E.~Goudzovski,
D.~Gurev, 
P.~Hristov\footnotemark[1],
V.~Kekelidze,
%     V.~Kozhuharov,
L.~Litov,
D.~Madigozhin,
N.~Molokanova,
Yu.~Potrebenikov,
S.~Stoynev,
A.~Zinchenko\\
{\em \small Joint Institute for Nuclear Research, Dubna, Russian    Federation} \\[0.2cm] 
 %no list received!
E.~Monnier\footnotemark[8],
E.~Swallow,
R.~Winston\\
{\em \small The Enrico fermi Institute, The University of Chicago, Chicago, Illinois, 60126, U.S.A.}\\[0.2cm]
 R.~Sacco\footnotemark[9],
 A.~Walker \\
{\em \small Department of Physics and Astronomy, University of    Edinburgh, JCMB King's Buildings, Mayfield Road, Edinburgh,    EH9 3JZ, U.K.} \\[0.2cm] 
 %no list received!
%
W.~Baldini,
P.~Dalpiaz,
P.L.~Frabetti,
A.~Gianoli,
M.~Martini,
F.~Petrucci,
M.~Scarpa,
M.~Savri\'e \\
{\em \small Dipartimento di Fisica dell'Universit\`a e Sezione    dell'INFN di Ferrara, I-44100 Ferrara, Italy} \\[0.2cm] 
 %no list received!
%
%
A.~Bizzeti\footnotemark[10],
M.~Calvetti,
G.~Collazuol\footnotemark[11],
G.~Graziani,
E.~Iacopini,
M.~Lenti,
F.~Martelli\footnotemark[12],
G.~Ruggiero\footnotemark[1],
M.~Veltri\footnotemark[12] \\
{\em \small Dipartimento di Fisica dell'Universit\`a e Sezione    dell'INFN di Firenze, I-50125 Firenze, Italy} \\[0.2cm] 
%no list received!
%
%
M.~Behler,
K.~Eppard,
 M.~Eppard\footnotemark[1],
 A.~Hirstius\footnotemark[1],
 K.~Kleinknecht,
 U.~Koch,
L.~Masetti, 
P.~Marouelli,
U.~Moosbrugger,
C.~Morales Morales,
 A.~Peters\footnotemark[1],
 M.~Wache,
 R.~Wanke,
 A.~Winhart \\
{\em \small Institut f\"ur Physik, Universit\"at Mainz, D-55099 Mainz,
Germany\footnotemark[13]} \\[0.2cm] 
A.~Dabrowski,
T.~Fonseca Martin,
M.~Velasco \\
{\em \small Department of Physics and Astronomy, Northwestern University, Evanston Illinois 60208-3112, U.S.A.}
 \\[0.2cm] 
G.~Anzivino,
P.~Cenci,
E.~Imbergamo,
G.~Lamanna,
P.~Lubrano,
A.~Michetti,
A.~Nappi,
M.~Pepe,
M.C.~Petrucci,
M.~Piccini,
M.~Valdata \\
{\em \small Dipartimento di Fisica dell'Universit\`a e Sezione    dell'INFN di Perugia, I-06100 Perugia, Italy} \\[0.2cm] 
 %no list received  !  
% 
%
 C.~Cerri,
F.~Costantini,
 R.~Fantechi,
 L.~Fiorini,
 S.~Giudici,
 I.~Mannelli,
G.~Pierazzini,
 M.~Sozzi \\
{\em \small Dipartimento di Fisica, Scuola Normale Superiore e Sezione dell'INFN di Pisa, I-56100 Pisa, Italy} \\[0.2cm] 
%no list received  !  
% 
%
C.~Cheshkov\footnotemark[1],
J.B.~Cheze,
 M.~De Beer,
 P.~Debu,
G.~Gouge,
G.~Marel,
E.~Mazzucato,
 B.~Peyaud,
 B.~Vallage \\
{\em \small DSM/DAPNIA - CEA Saclay, F-91191 Gif-sur-Yvette, France} \\[0.2cm] 
M.~Holder,
 A.~Maier,
 M.~Ziolkowski \\
{\em \small Fachbereich Physik, Universit\"at Siegen, D-57068 Siegen,
Germany\footnotemark[14]} \\[0.2cm] 
C.~Biino,
N.~Cartiglia,
M.~Clemencic, 
S.~Goy Lopez,
E.~Menichetti,
N.~Pastrone \\
{\em \small Dipartimento di Fisica Sperimentale dell'Universit\`a e    Sezione dell'INFN di Torino,  I-10125 Torino, Italy} \\[0.2cm] 
 W.~Wislicki,
\\
{\em \small Soltan Institute for Nuclear Studies, Laboratory for High    Energy
Physics,  PL-00-681 Warsaw, Poland\footnotemark[15]} \\[0.2cm] 
H.~Dibon,
M.~Jeitler,
M.~Markytan,
G.~Neuhofer,
L.~Widhalm \\
{\em \small \"Osterreichische Akademie der Wissenschaften, Institut  f\"ur
Hochenergiephysik,  A-1050 Wien, Austria\footnotemark[16]} \\[1cm] 

{\it Submitted to Physics Letters B}

\end{center}

\setcounter{footnote}{0}
\footnotetext[1]{Present address: CERN, CH-1211 Gen\`eve 23, Switzerland}
\footnotetext[2]{ Funded by the U.K.    Particle Physics and Astronomy Research Council}
\footnotetext[3]{Present address: Istituto di Cosmogeofisica del CNR di Torino, I-10133 Torino, Italy}
\footnotetext[4]{On leave from Sezione dell'INFN di Torino,  I-10125 Torino, Italy}
\footnotetext[5]{ On leave from \"Osterreichische Akademie der Wissenschaften, Institut  f\"ur Hochenergiephysik,  A-1050 Wien, Austria}
\footnotetext[6]{On leave from University of Richmond, Richmond, VA, 23173, 
USA; supported in part by the US NSF under award \#0140230}
\footnotetext[7]{Also at Dipartimento di Fisica dell'Universit\`a e Sezione dell'INFN di Ferrara, I-44100 Ferrara, Italy}
\footnotetext[8]{Also at Centre de Physique des Particules de Marseille, IN2P3-CNRS, Universit\'e 
de la M\'editerran\'e, Marseille, France}
% \footnotetext[9]{Present address: Laboratoire de l'Acc\'elerateur Lin\'eaire, IN2P3-CNRS, Universit\'e de Paris-Sud, 91898 Orsay, France}
\footnotetext[9]{Present address: Department of Physics, Queen Mary College, University of London,
Mile End Road, London E1 4NS, UK}
\footnotetext[10]{ Dipartimento di Fisica dell'Universita' di Modena e Reggio Emilia, via G. Campi 213/A I-41100, Modena, Italy}
\footnotetext[11]{Present address: Scuola Normale Superiore e Sezione dell'INFN di Pisa, I-56100 Pisa, Italy}
\footnotetext[12]{ Istituto di Fisica, Universita' di Urbino, I-61029  Urbino, Italy}
\footnotetext[13]{ Funded by the German Federal Minister for    Research and Technology (BMBF) under contract 7MZ18P(4)-TP2}
\footnotetext[14]{ Funded by the German Federal Minister for Research and Technology (BMBF) under contract 056SI74}
\footnotetext[15]{Supported by the Committee for Scientific Research grants
5P03B10120, SPUB-M/CERN/P03/DZ210/2000 and SPB/CERN/P03/DZ146/2002}
\footnotetext[16]{Funded by the Austrian Ministry for Traffic and 
Research under % the 
contract GZ 616.360/2-IV GZ 616.363/2-VIII, 
and by the Fonds f\"ur   Wissenschaft und Forschung FWF Nr.~P08929-PHY}

\end{Authlist}

\end{center}

\end{titlepage}

\vspace{2cm}

%\begin{abstract}
{\bf{Abstract}}

{\small
The NA48 collaboration has measured the amplitude of the CP-conserving 
component of the decay $K_S^0 \rightarrow \pi^+ \pi^- \pi^0$ relative
to $K_L^0 \rightarrow \pi^+ \pi^- \pi^0$. For the characteristic 
parameter $\lambda$,
% The CP-conserving component of the decay $K_S^0 \rightarrow \pi^+ \pi^- \pi^0$
%  has been measured by the NA48 collaboration at CERN. For the parameter $\lambda$,
%  which characterizes this decay, 
the values
$Re~\lambda =  0.038 \pm 0.010$ and $Im~\lambda = -0.013 \pm 0.007$ have
been extracted. These values agree with earlier measurements and with 
theoretical predictions from chiral perturbation theory.
}
%\end{abstract}

%\tableofcontents

\section{Introduction}

The $K_S^0 \rightarrow \pi^+ \pi^- \pi^0$ decay amplitude is dominated by two
% can be decomposed into 
angular momentum components, $l=0$           %(CP=-1) 
and $l=1$                                   % (CP=+1);
($l$ is the angular momentum of the neutral pion with respect to the system
of the two charged pions);
higher angular momentum states are suppressed because the kaon mass is
close to the three pion mass.  In this analysis we have measured the
CP conserving transition to the 
$l=1$ state through its interference with the dominant 
$K_L^0 \rightarrow \pi^+ \pi^- \pi^0$ decay.
We neglect any effects from CP violation
in mixing and decay, which are totally negligeable at our level of
sensitivity.
 
%The $K_S^0 \rightarrow \pi^+ \pi^- \pi^0$ decay amplitude has two
%angular momentum components, $l=0$ (suppressed by CP violation) and
%$l=1$ (CP conserving). In this analysis we have measured the $l=1$ state 
%by observing the interference with the dominant 
%$K_L^0 \rightarrow \pi^+ \pi^- \pi^0$ decay. 

% While the decay $K_L^0 \rightarrow \pi^+ \pi^-$ always violates CP, the 
% decay $K_S^0 \rightarrow \pi^+ \pi^- \pi^0$ has a CP-violating and a 
% CP-conserving component. This is explained by the different angular 
% momentum states $L$ in which the three decay particles may be
% ($l=0$ or $l=1$; higher angular 
% momenta are strongly suppressed). The CP-violating decay has $l=0$ while
% for the CP-conserving decay  $l=1$. So, while one channel is suppressed
% by CP-violation, the other is suppressed by angular 
% momentum.

The amplitudes for the decays of neutral kaons into three pions
($A_L^{3\pi}$ for $K_L^0$ and  $A_S^{3\pi}$ for $K_S^0$)
can be parameterized in terms of the Dalitz variables $X$ and $Y$, which
are defined as 
\begin{equation}
X = \frac { s_{\pi^-} - s_{\pi^+}}  { m_{\pi^{\pm}}^2 } ,
%\end{equation} 
%and 
%\begin{equation}
\ \ \ \ \ \ \ \ \ \ \ \ \ \ \ 
Y = \frac { s_{\pi^0} - s_0}  { m_{\pi^{\pm}}^2 }
\end{equation} 
with 
%\begin{equation}
$
%        s_{\pi} =  (p_K - p_{\pi})_{\mu} (p_K - p_{\pi})^{\mu},
s_{\pi} =  (p_K - p_{\pi})^2,
$
%\end{equation} 
$
s_0 =  \frac{1}{3} (s_{\pi ^+} + s_{\pi ^-} + s_{\pi ^0} ), 
$
and $p_K$ and $p_{\pi}$ being the 4-momenta of the kaon and the pion
respectively. The Dalitz variable $X$ is a measure of the difference of
the energies of the two charged pions in the kaon's rest system
while $Y$ is a measure of the energy of the neutral pion in the kaon's 
rest system.

%  The CP-conserving and the CP-violating component of the decay
The $l=1$ and the $l=0$ components of the decay
$K_S^0 \rightarrow \pi^+ \pi^- \pi^0$
can be separated by the fact that the amplitude of the
%CP-conserving 
$l=1$ process is antisymmetric in $X$
while the 
%CP-violating 
$l=0$ process is symmetric in $X$. 
% contains only quadratic terms in $X$ (and is thus even in $X$). 
%%When integrating over the whole range of $X$, the $l=1$
%CP-conserving 
%%term cancels, and the 
%CP-violating 
%%$l=0$ term can be measured. 
%%Conversely, the 
%CP-conserving 
Therefore the
$l=1$ contribution can be extracted by separately integrating 
over the regions $X>0$ and $X<0$ and subtracting the results from each 
other. We consider the distribution 
\begin{equation}
V(t) =  \frac { N_{3\pi}^{X>0}(t)  -  N_{3\pi}^{X<0}(t) }
      { N_{3\pi}^{X>0}(t)  +  N_{3\pi}^{X<0}(t) }          
\label{V_exp}  
\end{equation}
where $N_{3\pi}^{X>0}(t)$     [$N_{3\pi}^{X<0}(t)$]   is the number of
% $K^0 \rightarrow \pi^+ \pi^- \pi^0$ decays 
% reconstructed 
%at kaon decay 
decays of neutral kaons into $\pi^+ \pi^- \pi^0$ at
time $t$ with a value of the Dalitz variable $X$ larger  
[smaller] than zero~\cite{AmbrosPaver,AmbrosIsi}.

% Most $\pi^+ \pi^- \pi^0$ decays in the detector are due to the dominating 
% $K_L^0$-decay, and it is only via its interference with this decay that the
% rare decay $K_S^0 \rightarrow \pi^+ \pi^- \pi^0$ can be measured.
% For the $l=1$ 
%(CP-conserving) 
% decay, this interference can be described by 
% a complex parameter defined as
In the NA48 set-up, where $K_S^0$ and $K_L^0$ are produced in equal amounts at
a fixed target, the interference of the $l=1$ component of 
$K_S^0 \rightarrow \pi^+ \pi^- \pi^0$ with the $l=0$ component of the dominant 
$K_L^0 \rightarrow \pi^+ \pi^- \pi^0$ decay can be observed. 
It can be described by a complex 
parameter~\cite{CPLEAR98}

%\begin{equation}
%\lambda = \frac { \int_{Y_{min}}^{Y_{max}}\int_0^{X_{lim}(Y)}
%A_L^{* \  3\pi(CP=-1)}(X,Y)   A_S^{3\pi(CP=+1)}(X,Y) \  dXdY }
%{\int_{Y_{min}}^{Y_{max}}\int_0^{X_{lim}(Y)}
%| A_L^{* 3\pi(CP=-1)}(X,Y)|^2 \  dXdY}               ,
%\label{lambdadef}
%\end{equation}

\begin{equation}
\lambda = \frac { \int_{-\infty}^{\infty} dY \int_0^{\infty} dX \ 
%                             A_L^{* \  3\pi(CP=-1)}(X,Y)   A_S^{3\pi(CP=+1)}(X,Y)}
A_L^{* \  3\pi(l=0)}(X,Y)   A_S^{3\pi(l=1)}(X,Y)}
{\int_{-\infty}^{\infty} dY \int_0^{\infty} dX \ 
%                             | A_L^{* 3\pi(CP=-1)}(X,Y)|^2}               ,
| A_L^{3\pi(l=0)}(X,Y)|^2}               ,
\label{lambdadef}
\end{equation}
which has been extracted by fitting the distribution defined above:
%  from the $K^0 \rightarrow \pi^+ \pi^- \pi^0$
%  decay distribution over the kaon lifetime by separating the integration
%  over the two halves of the Dalitz plot in $X$ 
% \cite{CPLEAR98}:

\begin{equation}
V(t)  \approx 
\frac { 2 D(E) [Re(\lambda )cos( \Delta m t) - Im( \lambda )sin( \Delta m t)]
             e^{ - \frac{t}{2} (\frac {1}{\tau_S} + \frac {1}{\tau_L}   ) } }
      { e^{-\frac{t}{\tau_L}} } .
\label{V_formula}
\end{equation}
$\Delta m$ is the mass difference between $K_L^0$ and $K_S^0$, $\tau _L$ and
$\tau _S$ are the respective lifetimes, and the energy-dependent 
``dilution'' $D(E)$ is the difference in the relative abundances of 
$K^0$ and $\overline{K}^0$ at production for a kaon energy of $E$:
\begin{equation}
D(E) = \frac {K^0 - \overline{K}^0 } 
            {K^0 + \overline{K}^0 }        
\label{dilution_definition}
\end{equation}

% Using results from Chiral Perturbation Theory \cite{AmbrosIsi} one can 
% calculate a theoretical value of Re $\lambda$ = 0.031 and Im $\lambda$ = -0.006 
% \cite{CPLEAR98}.

\section{Experimental setup}

The data presented here were taken in a high-intensity neutral beam at
the CERN Super Proton Synchrotron during 89 days in 2002.
% by the NA48 experiment,
% a fixed-target experiment at the CERN Super Proton Synchrotron 
%(Geneva,Switzerland)
% in 2002.
Protons of 400 GeV with an average 
intensity of  $5 \times 10^{10}$  per 4.8~s spill
(within a cycle of 16.2~s) 
% from the SPS 
impinged on a beryllium 
target, where secondary particles were produced. 
Charged particles were removed by a sweeping magnet and a
5.1~m long collimator, which selected a beam of neutral particles at an angle of 4.2 mrad with
respect to the proton beam. 

After the collimator, the neutral beam and the decay products
propagated in an 89~m long vacuum tank, where most of the 
short-lived kaons ($K_S^0$) and 
neutral hyperons ($\Lambda^0, \Xi^0$) as well as a small fraction
of the long-lived neutral kaons ($K_L^0$) decayed. 
Undecayed particles ($K_L^0$, neutrons) continued to a beam dump via a vacuum 
pipe that passed through all of the downstream detectors. 

Downstream of the vacuum tank there was a magnetic spectrometer consisting of
four drift chambers inside a helium-filled tank and a magnet with a horizontal
transverse momentum kick of 265 MeV/c.
The spatial resolution was 120 $\mu$m per view, and the momentum resolution was 
$\delta p / p = ( 0.48 \oplus 0.015 \ p ) \%$, with momentum $p$ in GeV/c.

After the spectrometer there was a scintillator hodoscope, for the accurate timing 
of charged particles (yielding a time resolution on single tracks of 250~ps),
and a liquid-krypton electromagnetic calorimeter, for the measurement
of photons and electrons, with an energy resolution of
$\sigma(E)/E = ( 3.2/\sqrt{E} \oplus 9/E \oplus 0.42 )\% $ ($E$ in GeV).
These were followed by a hadron calorimeter and a muon detector. 
A more detailed description of the detector has been given elsewhere~\cite{pi0ee}.

\section{The trigger}

Decays into $\pi^+ \pi^- \pi^0$ made up only a small part of the event rate in
the detector. This was due to the small expected branching ratio for the $K_S^0$  decay
and to the long lifetime of the $K_L^0$.
%, of which only a small part decayed in the 
%                     fiducial volume of the detector.
To obtain a statistically significant sample of $\pi^+ \pi^- \pi^0$ decays, a very
selective trigger was used
% is needed 
to suppress the high two-particle background, which 
came mostly from  $K_S^0 \rightarrow \pi \pi$ and from 
$\Lambda^0 \rightarrow p \pi^-$.

When the level-1 trigger indicated an event with at least two charged tracks
and a minimum energy deposition of 30 GeV in the calorimeters, 
a fast online processor calculated the momenta of
the charged tracks, and hence the invariant mass of the decaying neutral
particle under the assumptions of 
$K_S^0 \rightarrow \pi^+ \pi^-$, $\Lambda^0 \rightarrow p \pi^-$, 
and $\overline{\Lambda}^0 \rightarrow 
\overline{p} \pi^+$. Events were accepted by the trigger if none 
of the calculated masses matched the physical mass of the particle in question
(difference of squares of the calculated mass and the nominal mass 
\cite{PDG2004} larger than two percent).
Additional cuts were applied to the relative momenta of the two 
charged particles ($p_{larger} / p_{smaller} < 3.5$, to further 
suppress $\Lambda^0$ decays) 
and to the separation of the two tracks at the entry of the spectrometer 
($>5$ cm in drift chamber~1, to remove photon conversions in the 
upstream Kevlar window).
To allow a measurement of the trigger efficiency, a downscaled sample 
of minimum-bias events (1/35) was recorded in parallel, 
where only the level-1 trigger condition was fulfilled. 
In the course of data-taking, the high voltage in the drift chambers
was occasionally adjusted to maintain stable operating conditions. 
As a result, the efficiency of the level-2 trigger varied between   
(75$\pm$4)\% and (86$\pm$4)\%. By  comparing the  analysis of
different time periods , it was checked that this variation did
not introduce any spurious asymmetries.

% During the course of the data taking, the high voltage in the drift 
% chambers had to be gradually reduced because of technical problems. As a 
% result, the level-2 trigger efficiency went down from (86$\pm$4)\% to 
% (75$\pm$4)\%. 
% It has been verified that this global change in the trigger efficiency did 
% not create any spurious asymmetries and so did not affect the present 
% analysis. 

\section{Data analysis}

For the offline analysis, events were selected with  
two tracks of opposite charge from the same vertex (distance of closest 
approach  less than 3~cm), and at least two electromagnetic clusters 
(within $\pm 20$~ns).
For each pair of clusters, the $\pi^0$-mass was calculated assuming that the 
% kaon 
pion decayed at the position of the charged vertex. If there were more than
two clusters, the cluster pair yielding the best $\pi^0$-mass was retained.
 
In order to reduce acceptance effects near the edges of the detectors, each charged track was required to cross the drift chambers at a minimum
distance of 12~cm and a maximum distance of 110~cm from the beam axis. 
Likewise, each photon cluster was required to be more than 15~cm from the beam axis and more than 10~cm from the outer edge of the calorimeter.

In addition, the photon clusters had to be well separated from tracks ($>$15~cm) to avoid mismeasurements due to energy sharing. 
Tracks were identified as charged pions rather than electrons by demanding
that their electromagnetic energy deposition was less than
90 percent of the track momentum. 
Cuts were applied on the charged vertex (less than 3~cm lateral distance from
the $K^0$ beam line, and between 800 and 5400~cm from the target), 
and on the masses reconstructed for 
$K^0 \rightarrow \pi^+ \pi^- \pi^0$ and $\pi^0 \rightarrow \gamma \gamma$ 
($m_{K^0}^{PDG} \pm 10.5$~MeV and $m_{\pi^0}^{PDG} \pm 7.8$~MeV~\cite{PDG2004}). 
These cuts also excluded $K_S^0 \rightarrow \pi^+ \pi^-$ decays. 
The reconstructed kaon energy was required to be between 30 and 166~GeV.
$\Lambda^0$ hyperons were further suppressed by cutting harder than
in the trigger on the masses for the hypotheses 
$\Lambda^0 \rightarrow p \pi^-$ and $\overline{\Lambda}^0 \rightarrow 
\overline{p} \pi^+$
($m_{\Lambda^0}^{PDG} \pm 21$~MeV~\cite{PDG2004})
and on the momentum ratio       % difference 
of the two charged particles 
($p_{larger} / p_{smaller} < 2.9$). No $\Lambda^0$ hyperons survive after
these two cuts. The mass cuts were optimized
to maintain a high efficiency without introducing substantial background. 
A total of 19 million events passed all these cuts. 

\begin{figure}[ht]
\vspace{8cm}
\includegraphics{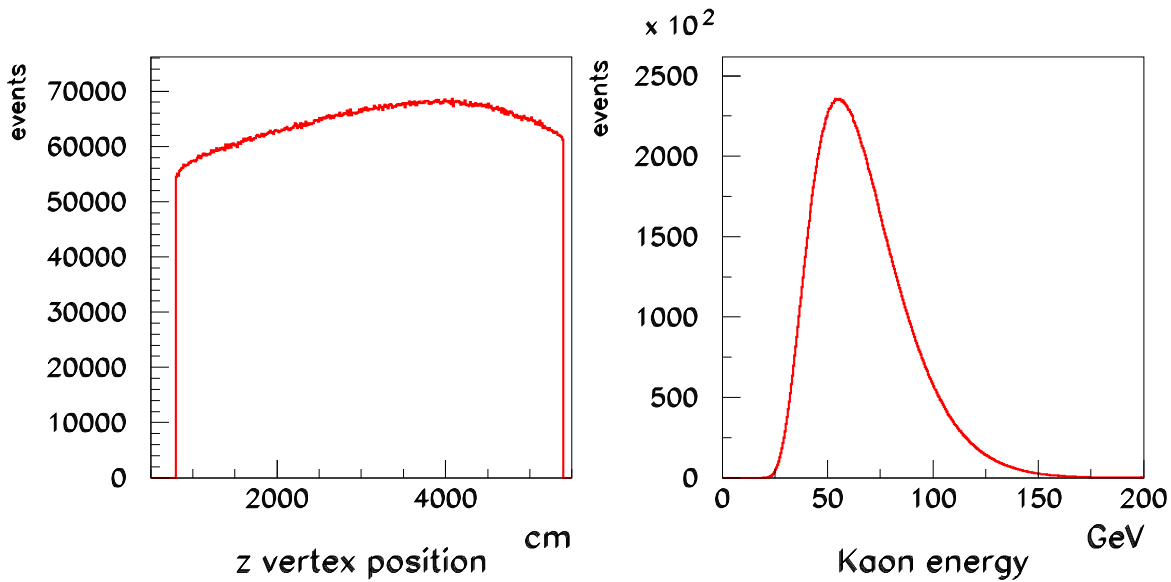}
\caption{Position of the decay vertex along the beam line (left,
measured from target position)
and reconstructed $K^0$ energy (right) for the data sample of 19 million events.
\label{E_z_distr}}
\end{figure}

Fig.~\ref{E_z_distr} shows the distributions of the position of decay
vertices along the beam line (z vertex) and the reconstructed $K^0$ energy
after applying the analysis cuts.

\begin{figure}[ht]
\vspace{8cm}
\includegraphics{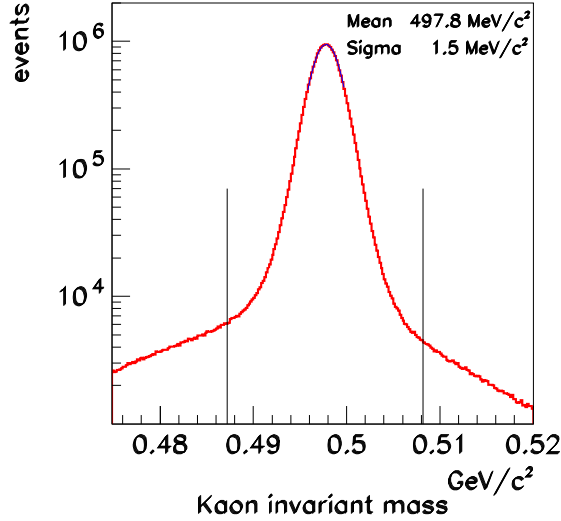}
\caption{The $K^0$ invariant mass. The vertical lines indicate the 
$K^0$ mass cut used in the analysis.
% with the cuts described in the text.
% The mean and sigma were obtained by a gaussian fit around the peak.
\label{massK}}
\end{figure}

Fig.~\ref{massK} shows the distribution of the $K^0$ invariant mass 
%and for the $\pi^0$ invariant mass 
after cuts.

\begin{figure}[h]
\vspace{14cm}
\includegraphics{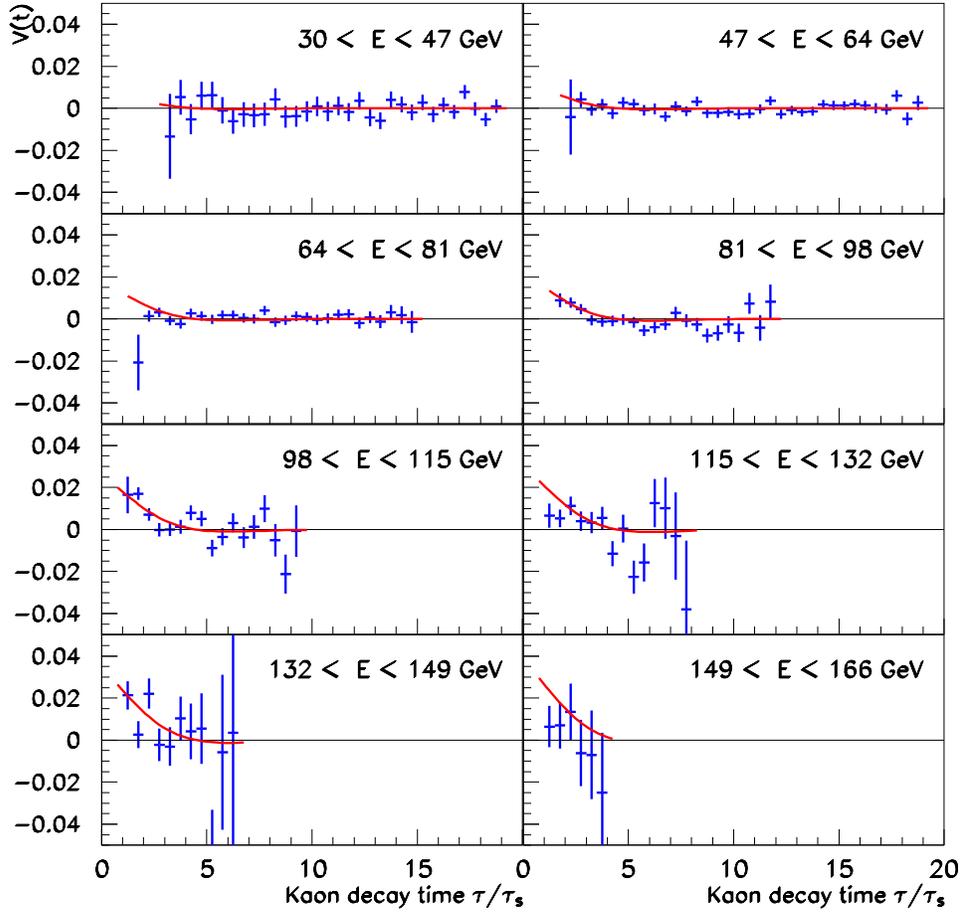}
\caption{The experimental distribution $V(t)$ (points with error bars) and the 
function obtained by a $\chi^2$ fit (smooth curve) in bins of time
and energy. The $\chi^2$ per degree of freedom is 169.35/162=1.045.
The fit was made simultaneously over all energy bins, but is
mainly constrained by the intermediate energy bins. 
% The highest energy bins contribute little because of the large 
% statistical errors
% while in the lowest energy bins the interference effect dies out before
% the particles leave the collimator. 
The data were compensated for acceptance and trigger efficiency by applying
Monte-Carlo corrections and by averaging over samples with opposite 
magnetic field in the spectrometer. 
\label{bindatafit}}
\end{figure}

Applying these cuts and correcting for acceptance and trigger efficiency 
as described below, the experimental distribution $V(t)$ defined in 
(\ref{V_exp}) was obtained. 
% Using the                                   %literature 
% values for $\Delta m$, $\tau _L$ and $\tau _S$ 
% from \cite{PDG2004} and for $D(E)$ from \cite{Car90} (adapted to the
% different proton energy and production angle by using measurements  
% with charged kaons \cite{AtherDoble}), t
The values of $Re~\lambda$
and $Im~\lambda$ were extracted by fitting equation (\ref{V_formula}) to the 
experimental data. Due to the energy dependence of the dilution $D(E)$,
the fit was made in bins of kaon energy (see Fig.~\ref{bindatafit}), 
with 8 equal-sized energy bins ranging from 30 to 166 GeV. 
The values for the external parameters $\Delta m$, $\tau _L$ and $\tau _S$
were taken from \cite{PDG2004}. For the dilution $D(E)$, we used a
quadratic fit to the values measured in  \cite{Car90} (energy $E$ given
in GeV): 
\begin{equation}
D_{measured}(E) = -0.167 + \frac{6.10}{10^3}  E -  \frac{1.72}{10^5} E^2
% D_{measured}(E) = -0.167 + 6.10 \times 10^{-3} \times E -
% 1.72 \times 10^{-5} \times  E^2
\end{equation}
The errors for the individual points measured in \cite{Car90} vary between
5 percent at the lowest kaon energies (75 GeV) and 25 percent at the highest 
energies (165 GeV). We assume the uncertainties at the different energies
to be uncorrelated.
To take into account the difference in proton energy and production angle
between \cite{Car90} and our own measurement, we used measurements  
with charged kaons \cite{AtherDoble}, from which we derived an empirical
correction factor (energy $E$ given in GeV): 
\begin{equation}
D_{corr}(E) = 1.28 
- \frac{3.82}{10^3} E 
+ \frac{5.38}{10^5}  E^2
- \frac{3.32}{10^7}  E^3
+ \frac{7.15}{10^{10}} E^4
% - 3.82 \times 10^{-3} \times E 
% + 5.38 \times 10^{-5} \times E^2
% - 3.32 \times 10^{-7} \times E^3
% + 7.15 \times 10^{-10} \times E^4
\end{equation}
To fit equation (\ref{V_formula}), we then use 
\begin{equation}
D(E) = D_{measured}(E) \times D_{corr}(E)
\end{equation}
Based on the uncertainties in \cite{Car90} and \cite{AtherDoble}, the
overall relative uncertainty in $D(E)$ is estimated at about 15 percent.

% \clearpage

The aim of the experiment was to measure a small effect that is antisymmetric
in the Dalitz variable $X$ (the interference of $K_S^0$ and $K_L^0$ 
in the decay
into $\pi^+ \pi^- \pi^0$) over a large background that is symmetric in $X$
(from the CP-conserving decay $K_L^0 \rightarrow \pi^+ \pi^- \pi^0$). 
Therefore, care had to be taken to
exclude any detector effects that could artificially introduce such an 
asymmetry in the $K_L^0$ decay. 

The sign of $X$ equals the sign of the momentum difference of the $\pi^+$ and
the $\pi^-$ in the kaon rest system. The pion momenta in the kaon rest system
are correlated to their momenta in the laboratory system. Any momentum-dependent
difference in the acceptance or the trigger efficiency 
for positive and negative pions may therefore 
introduce a possible bias. Due to the field of the spectrometer magnet,
positive and negative particles populated the detector in a different way.
In order to compensate such effects, the field of the spectrometer
magnet was reversed weekly during the data taking. 

An artificial asymmetry can be caused by the detector acceptance only
if this acceptance is asymmetric in $X$. However, if there is a real, physical
asymmetry, its value may be distorted even by a symmetric $X$-dependence
of the acceptance. 
%When excluding the trigger inefficiency described above, 
The               % Monte Carlo 
% for the 
NA48 detector simulation (based on the GEANT3 package \cite{GEANT})
showed a symmetric decrease in the acceptance for 
high $|X|$. Using the simulation, the acceptance was calculated
as a function of $|X|$, kaon decay time and kaon energy. The effect was 
then corrected by weighting each event in the data in inverse 
proportion to the calculated acceptance for the corresponding value of 
$|X|$, kaon decay time and kaon energy.
% high $|X|$, which was taken into account by weighting each event in inverse 
% proportion to the acceptance calculated for the corresponding value of $|X|$.
This procedure resulted an increase of $10 \pm 5$ percent in the extracted
values of $Re~\lambda$ and $Im~\lambda$.

In the drift chambers, signal attenuation near the ends of some wires created 
a left-right acceptance asymmetry on the detector periphery, especially for 
strongly deviated low-momentum particles.  This was compensated in the 
offline reconstruction by the redundancy of chamber planes, and there was 
no visible effect on the minimum-bias data sample. 
However the data collected with the fast level-2 trigger algorithm, which 
did not use the full redundancy, showed an asymmetry in $X$ at all kaon 
lifetimes,
which was strongest (2 percent) at the lowest kaon energies (30 GeV).
This asymmetry was reversed when the magnetic field of the spectrometer
magnet was inverted.
Using the $X$-symmetric amplitude of $K_L \rightarrow \pi^+ \pi^- \pi^0$,
the detector and trigger simulation accurately reproduced this behavior. 

The Monte-Carlo distributions from the simulation were used to correct 
the data. 
The correction resulted in a shift of about $3 \times 10^{-2}$ 
in both $Re~\lambda$ and $Im~\lambda$, with opposite sign for the two
field orientations of the spectrometer magnet.
The simulation program had been developed, extensively used and validated
during previous studies by the collaboration \cite{pi0ee,EpsPrime}.
After acceptance and trigger corrections, the individual results 
for the two field orientations differ by about 1.4 $\sigma$.
To exclude possible additional effects 
% not included in the Monte Carlo, 
the Monte-Carlo corrected data samples with positive and negative 
fields in the spectrometer were subsequently averaged 
(taking the arithmetic mean 
of the observed distributions $V(t)$ in formula (\ref{V_exp}), which 
is equivalent to normalizing
the data taken in the two field orientations to the same number of events and
summing them before applying (\ref{V_exp})).
Slightly different results
were obtained when using either only the Monte-Carlo correction (without
averaging over the two magnetic field orientations), or only averaging
over the magnetic fields (without the Monte-Carlo correction). We use half the
difference between these two last approaches as a contribution to
our systematic error 
($\pm$0.004 in $Re~\lambda$ and 
$\pm$0.003 in $Im~\lambda$). 

\clearpage

Another cross-check was to analyse the data with a tight outer radius cut
of 50~cm in the drift chambers and otherwise follow the above procedure. 
This was prompted by the fact that the observed drift chamber
inefficiencies were highest at large radii. The results in $Re~\lambda$
and $Im~\lambda$ differed by about 0.001 while the statistical errors
increased by about the same amount.

The other major contribution to the systematic uncertainty comes from the
$K^0 / \overline{K}^0$ 
dilution $D(E)$ defined in formula~(\ref{dilution_definition}) and leads 
to an additional uncertainty of $\pm$0.005 in $Re~\lambda$ and of
$\pm$0.003 in $Im~\lambda$.
By varying the experimental cuts quoted above within reasonable bounds
and the external parameters from \cite{PDG2004} within their errors it was
seen that all other possible sources of systematic uncertainties 
were negligible.
% can be neglected at this level of uncertainty. 

% The remaining systematic error was estimated to be 
% $\pm$0.005 in both $Re~\lambda$ and $Im~\lambda$
% by varying the cuts 
% quoted above and the value used for the $K^0 / \overline{K}^0$ 
% dilution $D(E)$ defined in formula~(\ref{dilution_definition}) 
% within reasonable bounds. This variation was applied both to the experimental
% data and to the signal Monte Carlo. 

\begin{figure}[ht]
\vspace{8cm}
\includegraphics{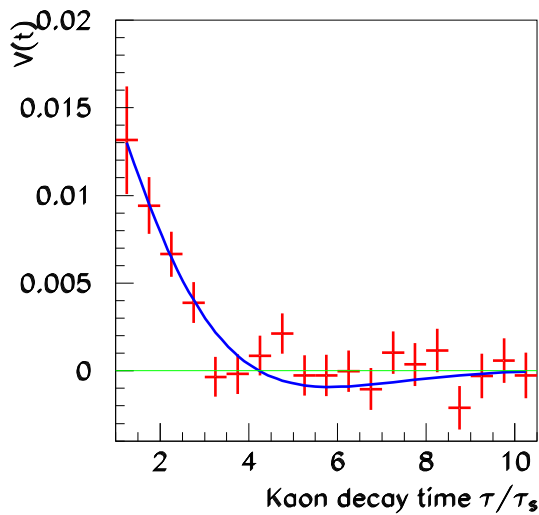}
\caption{The experimental distribution $V(t)$ (points with 
error bars, data summed over all kaon energies) and the function
obtained from the fitted values for $Re~\lambda$ and $Im~\lambda$ 
for a mean kaon energy of 87 GeV.
\label{alldatafit}}
\end{figure}

\section{Results}

The following values have been obtained for the real and imaginary parts
of $\lambda$: 
\begin{eqnarray}
Re~\lambda = +0.038 \pm 0.008_{stat} \pm 0.006_{syst}\\
Im~\lambda = -0.013 \pm 0.005_{stat} \pm 0.004_{syst}  
\end{eqnarray}

Adding both errors in quadrature, 
\begin{eqnarray}
  Re~\lambda = +0.038 \pm 0.010\\
  Im~\lambda = -0.013 \pm 0.007
\end{eqnarray}

These values can be compared with the theoretical numbers of 
$Re~\lambda$=+0.031 and $Im~\lambda$=-0.006
obtained in \cite{CPLEAR98} by using results from 
Chiral Perturbation Theory \cite{AmbrosIsi}.

Fig.~\ref{alldatafit} shows the distribution $V(t)$ defined in 
(\ref{V_exp}) 
% of the $X$-asymmetry over lifetime 
derived from the data,
and the curve obtained 
from (\ref{V_formula}) with the fitted values of $Re~\lambda$ and $Im~\lambda$
(central curve, plotted for a mean kaon energy of 87 GeV; the top and bottom
curves result from varying the dilution within errors).
%             and for an effective dilution of D=0.27. 
These values were not obtained by fitting this histogram directly,
but by fitting all data in bins of energy. This is important because of the
dependence of the dilution $D(E)$ on the kaon energy.

\begin{figure}[ht]
\vspace{8cm}
\includegraphics{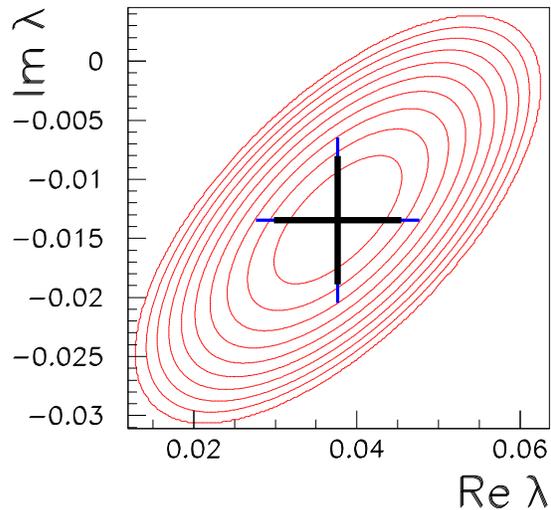}
\caption{The fitted values of $Re~\lambda$ and $Im~\lambda$ 
in the complex plane with the 
contours for a $\chi^2$ of 1 to 10 above the minimum $\chi^2$.
The  
shorter error bars and the contours refer only to the statistical 
uncertainty while the
longer error bars correspond to the total uncertainty.
%\caption{The fitted value of Re~$\lambda$ and Im~$\lambda$ 
%in the complex plane with the 
%contours for a $\chi^2$ of 1, 4, and 9 above the minimum $\chi^2$,
%corresponding to a confidence level of 39\%, 86\% and 99\% that the
%real value lies within the corresponding ellipse.
% 1$\sigma$, 2$\sigma$ and 3$\sigma$ errors in the
% complex plane. 
%Data taken with the minimum-bias trigger
%have not been included in this plot.
%Only the statistical error is shown.
\label{chi2}}
\end{figure}

The real and imaginary parts of $\lambda$ are strongly correlated, as can be
seen from Fig.~\ref{chi2}, which shows the measured values and 
% the $1\sigma$, $2\sigma$ and $3\sigma$ 
contours of equal $\chi^2$ in the complex plane. 
% Re $\lambda$ and Im $\lambda$ are strongly correlated (correlation coefficient = 0.66).
Taking into account only the statistical errors, 
the correlation coefficient between $Re~\lambda$ and $Im~\lambda$ is 0.66.

From the real part of $\lambda$, the branching ratio for the 
CP-conserving component of the decay 
$K_S^0 \rightarrow \pi^+ \pi^- \pi^0$ can be obtained. Using the same
approximation and the external parameters as in \cite{CPLEAR98}, we get:

\begin{center}
$BR(K_S^0 \rightarrow \pi^+ \pi^- \pi^0) = 
%               (4.7~ ^{+3.7}_{-2.7}) \times 10^{-7}$
(4.7~ ^{+2.2}_{-1.7} (stat) \   ^{+1.7}_{-1.5} (syst)    ) \times 10^{-7}$
\end{center}

% These results agree with two other recent measurements \cite{CPLEAR98,FermiE621}.
The results from this experiment agree with chiral perturbation theory and 
with two other measurements with comparable errors \cite{CPLEAR98,FermiE621}. 
It is worth noting that experiment \cite{CPLEAR98} used tagged kaons, so that
its result is independent of any measurements or calculations of the 
dilution $D(E)$ while \cite{FermiE621} was carried out at higher kaon energies
and therefore higher dilution than our measurement. 
% It should also be noted that when combining \cite{CPLEAR98} and \cite{FermiE621}
% with our own measurement, the imaginary part of $\lambda$ (which is predicted
% to be small) differs significantly from zero.
It should also be noted that all three experiments obtain a non-zero negative 
value for the imaginary part of $\lambda$ (which is predicted
to be small).\\
\\
\\

% \clearpage

{\bf Acknowledgements}

It is a pleasure to thank the technical staff of the participating laboratories,
universities and affiliated computing centers for their efforts in the construction
of the NA48 apparatus, in the operation of the experiment, and in the processing
of the data. We would also like to thank Maria Fidecaro for useful discussions.

\end{document}